\documentclass[8pt]{article}
\usepackage{graphicx}
\usepackage{amssymb}
\def\beq{\begin{eqnarray}}
\def\eeq{\end{eqnarray}}

\usepackage{amsfonts}

\begin{document}


\title{Perturbation theory for short-range weakly-attractive potentials in one dimension}
\author{Paolo Amore \\
\small Facultad de Ciencias, CUICBAS, Universidad de Colima,\\
\small Bernal D\'{i}az del Castillo 340, Colima, Colima, Mexico \\
\small paolo.amore@gmail.com \\
Francisco M. Fern\'andez \\
\small INIFTA (UNLP, CONICET), Division Qu\'imica Te\'orica, \\
\small  Blvd. 113 S/N, Sucursal 4, Casilla de Correo 16, 1900 La Plata, Argentina \\
\small fernande@quimica.unlp.edu.ar}

\maketitle

\begin{abstract}
We have obtained the perturbative expressions up to sixth order for the energy of the bound state in
a one dimensional, arbitrarily weak, short range finite well, applying a method originally developed
by Gat and Rosenstein~Ref.~\cite{Gat93}. The expressions up to fifth order reproduce the results
already known in the literature, while the sixth order had not been calculated before. As an illustration
of our formulas we have applied them to two exactly solvable problems and to a nontrivial problem.
\end{abstract}


\section{Introduction}
\label{sec:Intro}

We consider the Schr\"odinger equation in one dimension
\begin{eqnarray}
\hat{H} \psi(x) = E \psi(x)
\label{eq_sch}
\end{eqnarray}
with
\begin{eqnarray}
\hat{H} = - \frac{d^2}{dx^2} + \lambda V(x)
\end{eqnarray}
where $V(x)$ is a potential of finite depth ($\lim_{|x|\rightarrow \infty} V(x) = 0$ and $V(x) < 0$ for $x \in (-\infty,\infty)$).

For this problem Simon\cite{Simon76} has stated the necessary and
sufficient conditions for a bound state to exist for  $\lambda
\rightarrow 0$, proving the analyticity of the lowest energy
eigenvalue at $\lambda=0$, in one dimension (in two dimensions, on
the other hand, Simon has also proved the non-analyticity of the
eigenvalue). The work of Simon was stimulated by the findings of
Abarbanel, Callan and Goldberger \cite{Abarbanel76}, who had
obtained the expression for the lowest eigenvalue to order
$\lambda^3$, when $V(x)$ is a short range potential.
Interestingly, as mentioned by Simon in a note added in proof, the
leading order term of this expansion had already been presented in
the ``Quantum Mechanics'' book by Landau and
Lifshitz~\cite{Landau}. More recently, Patil~\cite{Patil80a} has
obtained the perturbative expression for the lowest eigenvalue to
order $\lambda^5$ for short range potentials, using a perturbative
expansion for the inverse of the T matrix,  and discussed the case
of long range potentials as well.

Of particular interest to the present work, is the method
developed by Gat and Rosenstein in Ref.~\cite{Gat93}, which relies
on an appropriate modification of the unperturbed Hamiltonian, via
an attractive delta potential of arbitrarily small strength, which
allows one  to carry out the standard Rayleigh-Schr\"odinger
perturbation theory; the infrared divergences, which would be
present in the standard RS scheme, here identically cancel out and
the result is finite when, at the end of the calculation, the
strength of the delta potential is sent to zero. In this way, Gat
and Rosenstein reproduced the results of Abarbanel et
al.\cite{Abarbanel76}, obtaining the correct expression for the
energy to order $\lambda^3$.

In the present work, we have extended the calculation of Ref.~\cite{Gat93} to order $\lambda^6$, reproducing
all the known results up to order $\lambda^5$, contained in Ref.~\cite{Patil80a}, and obtaining
the exact contribution of order $\lambda^6$, which had not been previously calculated.
This work is organized as follows: in section \ref{sec:PT} we describe the method of Gat and Rosenstein;
in section \ref{sec:PC} we work out the explicit expressions for the contributions to the energy of the ground state to fourth, fifth and sixth order in
perturbation theory; in section \ref{sec:App} we discuss three applications of the formulas obtained in this paper;
finally, in section \ref{concl} we state our conclusions. \ref{AppA} contains the explicit expressions
for the Green's functions, appearing in the perturbative expressions.

\section{The method}
\label{sec:PT}

The first step in the application of the method is the suitable modification of the  Hamiltonian, introducing
a weak attractive delta potential:
\begin{eqnarray}
\hat{H} = \hat{H}_0 + \lambda V(x)
\end{eqnarray}
where
\begin{eqnarray}
\hat{H}_0 \equiv \left(- \frac{d^2}{dx^2} - 2 \beta \delta(x) \right)
\end{eqnarray}
is the ``unperturbed hamiltonian''.

The eigenfunctions of $\hat{H}_0$ are
\begin{eqnarray}
\psi_0(x) &=& \sqrt{\beta} e^{-\beta |x|} \nonumber  \\
\psi_p^{(even)}(x) &=& \frac{\sqrt{2}}{\sqrt{p^2+\beta^2}}  \left[ p \cos (px) - \beta \sin (p|x|) \right] \nonumber  \\
\psi_p^{(odd)}(x) &=& \sqrt{2} \sin (p|x|)\nonumber  \ .
\end{eqnarray}
and the corresponding eigenvalues are
\begin{eqnarray}
\epsilon_0 &=&  -\beta^2 \nonumber \\
\epsilon_p^{(even)} &=& \epsilon_p^{(odd)} = p^2\nonumber
\end{eqnarray}

In what follows we will adopt Dirac notation to denote the eigenstates of $\hat{H}_0$:
\begin{eqnarray}
\psi_0(x) &\rightarrow& |0\rangle \nonumber  \\
\psi_p^{(even)}(x) &\rightarrow& |p^{(even)}\rangle  \nonumber \\
\psi_p^{(odd)}(x)  &\rightarrow& |p^{(odd)}\rangle \nonumber
\end{eqnarray}

Although the lowest orders of this expansion can be found in most books on Quantum Mechanics
(ref.~\cite{Landau}, for instance, reports the expressions up to fourth order), the higher
orders must be calculated explicitly.
We report the general expressions for the perturbative corrections to the energy of the
ground state of Eq.(\ref{eq_sch}) up to sixth order, obtained using the NCAlgebra package
for Mathematica~\cite{ncalgebra}:
\begin{eqnarray}
E_0^{(1)} &=&  \langle 0 | V | 0 \rangle \nonumber \\
E_0^{(2)} &=&  -\langle 0 | V \hat{\Omega} V | 0 \rangle \nonumber \\
E_0^{(3)} &=&  \langle 0 | V \hat{\Omega} V \hat{\Omega} V | 0 \rangle
- \langle 0 | V | 0 \rangle \langle 0 | V \hat{\Omega}^2 V | 0 \rangle  \nonumber \\
E_0^{(4)} &=& \langle 0 | V \hat{\Omega} V  | 0 \rangle \langle 0 | V \hat{\Omega}^2 V | 0 \rangle
+2  \langle 0 | V | 0 \rangle \langle 0 | V \hat{\Omega}^2 V  \hat{\Omega} V | 0 \rangle \nonumber \\
&-& \langle 0 | V | 0 \rangle^2 \langle 0 | V \hat{\Omega}^3 V | 0 \rangle
-  \langle 0 | V \hat{\Omega} V \hat{\Omega} V \hat{\Omega} V | 0 \rangle \nonumber \\
E_0^{(5)} &=& -\langle 0 | V  | 0 \rangle^3 \langle 0 | V \hat{\Omega}^4 V  | 0 \rangle \nonumber \\
&+& \langle 0 | V  | 0 \rangle^2 \left(2 \langle 0 | V \hat{\Omega}^3 V \hat{\Omega} V | 0 \rangle + \langle 0 | V \hat{\Omega}^2 V \hat{\Omega}^2 V | 0 \rangle  \right)  \nonumber \\
&+& \langle 0 | V  | 0 \rangle \left( \langle 0 | V  \hat{\Omega}^2 V | 0 \rangle^2 - 2 \langle 0 | V  \hat{\Omega}^2 V \hat{\Omega} V\Omega V | 0 \rangle + 2 \langle 0 | V  \hat{\Omega}^3  V  | 0 \rangle \langle 0 | V  \hat{\Omega} V | 0 \rangle -
\langle 0 | V  \hat{\Omega} V \hat{\Omega}^2 V\Omega V | 0 \rangle \right) \nonumber \\
&+&  \left( - 2 \langle 0 | V  \hat{\Omega}^2 V \hat{\Omega} V | 0 \rangle  \langle 0 | V  \hat{\Omega} V  | 0 \rangle  -  \langle 0 | V  \hat{\Omega}^2 V  | 0 \rangle \langle 0 | V  \hat{\Omega} V \hat{\Omega} V| 0 \rangle + \langle 0 | V  \hat{\Omega} V \hat{\Omega} V\Omega V \hat{\Omega} V | 0 \rangle \right)
\nonumber\\
E_0^{(6)} &=&-\langle 0 | V  | 0 \rangle^4 \langle 0 | V \hat{\Omega}^5 V  | 0 \rangle \nonumber \\
&+& 2 \langle 0 | V  | 0 \rangle^3 \left(\langle 0 | V \hat{\Omega}^4 V \hat{\Omega} V | 0 \rangle +
  \langle 0 | V \hat{\Omega}^3 V \hat{\Omega}^2 V | 0 \rangle  \right)  \nonumber \\
&+&   \langle 0 | V  | 0 \rangle^2 \left( -2 \langle 0 | V \hat{\Omega}^3 V \hat{\Omega} V \hat{\Omega} V | 0 \rangle +
3\langle 0 | V \hat{\Omega}^3 V  | 0 \rangle \langle 0 | V \hat{\Omega}^2 V | 0 \rangle -
2 \langle 0 | V \hat{\Omega}^2 V \hat{\Omega}^2 V  \hat{\Omega} V | 0 \rangle \right. \nonumber \\
&-& \left.
\langle 0 | V \hat{\Omega}^2 V \hat{\Omega} V \hat{\Omega}^2 V| 0 \rangle
+ 3 \langle 0 | V \hat{\Omega}^4  V | 0 \rangle \langle 0 | V \hat{\Omega} V | 0 \rangle
- \langle 0 | V \hat{\Omega} V \hat{\Omega}^3 V \hat{\Omega} V | 0 \rangle
\right) \nonumber \\
&-& 2  \langle 0 | V  | 0 \rangle
\left( 2  \langle 0 | V \hat{\Omega}^2 V | 0 \rangle \langle 0 | V \hat{\Omega}^2 V \hat{\Omega} V | 0 \rangle
- \langle 0 | V \hat{\Omega}^2 V \hat{\Omega} V \hat{\Omega}  V \hat{\Omega} V| 0 \rangle
+ 2 \langle 0 | V \hat{\Omega}^3 V \hat{\Omega}  V| 0 \rangle \langle 0 | V \hat{\Omega} V| 0 \rangle   \right. \nonumber \\
&+& \left. \langle 0 | V \hat{\Omega}^2 V \hat{\Omega}^2 V| 0 \rangle \langle 0 | V \hat{\Omega} V | 0 \rangle
- \langle 0 | V \hat{\Omega} V \hat{\Omega}^2 V \hat{\Omega} V \hat{\Omega} V| 0 \rangle
+ \langle 0 | V \hat{\Omega}^3  V| 0 \rangle \langle 0 | V \hat{\Omega} V \hat{\Omega} V| 0 \rangle
\right) \nonumber \\
&+& \left( -2 \langle 0 | V \hat{\Omega}^2 V \hat{\Omega} V| 0 \rangle - \langle 0 | V \hat{\Omega}^2 V| 0 \rangle
\langle 0 | V \hat{\Omega} V \hat{\Omega} V| 0 \rangle   + \langle 0 | V \hat{\Omega} V \hat{\Omega} V \hat{\Omega} V \hat{\Omega} V| 0 \rangle  \right) \nonumber
\end{eqnarray}
where
\begin{eqnarray}
\hat{\Omega} \equiv \int_0^\infty \frac{dp}{2\pi}
\frac{|p^{(even)}\rangle \langle p^{(even)} |+|p^{(odd)}\rangle \langle p^{(odd)} |}{\epsilon_p-\epsilon_0} \ .
\end{eqnarray}

Upon using the explicit expressions for the Green's functions (reported in  \ref{AppA}) and taking
the limit $\beta \rightarrow 0^+$ at the end of the calculation, one can obtain the exact expressions for
the perturbative corrections to the energy of the ground state.

\section{Perturbative calculation}
\label{sec:PC}

The calculation of the corrections up to third order has been performed by Gat and Rosenstein in their paper \cite{Gat93},
therefore we concentrate on the next three orders. In this section we present the calculation of the fourth, fifth and
sixth orders, using the method of Gat and Rosenstein. The fourth and fifth orders obtained here reproduce the result
obtained earlier by Patil, whereas the sixth order is new.

\subsection{Fourth order}
\label{fourth}

The direct substitution of the expressions for the Green's functions inside the fourth order correction leads to a
rather lengthy expression; it is instructive to report this expression explicitly, that reads
\begin{eqnarray}
E_0^{(4)} &=& \lambda^4 \int dx_1 dx_2 dx_3 dx_4  V(x_1) V(x_2) V(x_3) V(x_4) \frac{\left| x_1-x_2\right| +\left| x_2-x_3\right| -2 \left|x_3-x_4\right| }{32 \beta }   \nonumber \\
&+& \lambda^4 \int dx_1 dx_2 dx_3 dx_4  V(x_1) V(x_2) V(x_3) V(x_4) \frac{\left(-3 x_1^2+6 x_2 x_1-8 x_2^2-x_3^2+4 x_4^2+10 x_2 x_3-8 x_3 x_4\right)}{64}  \nonumber \\
&+& \lambda^4 \int dx_1 dx_2 dx_3 dx_4  V(x_1) V(x_2) V(x_3) V(x_4) \mathcal{F}(x_1,x_2,x_3,x_4)
\end{eqnarray}
where
\begin{eqnarray}
 \mathcal{F}(x_1,x_2,x_3,x_4) &\equiv&  -\frac{5}{128} \left| x_1\right|  \left| x_1-x_2\right| -\frac{5}{128} \left| x_2\right|  \left| x_1-x_2\right| - \frac{1}{16} \left| x_2-x_3\right|  \left| x_1-x_2\right| \nonumber \\
&-& \frac{1}{16} \left| x_3\right|  \left| x_1-x_2\right|
   -\frac{1}{16} \left| x_3-x_4\right|  \left| x_1-x_2\right|
   -\frac{1}{16} \left| x_4\right|  \left| x_1-x_2\right| \nonumber \\
&+& \frac{x_2 \left| x_1\right|  \left| x_1-x_2\right| }{128
   x_1}+\frac{x_1 \left| x_2\right|  \left| x_1-x_2\right| }{128
   x_2}-\frac{1}{128} \left| x_1\right|  \left| x_1+x_2\right| \nonumber \\
&-& \frac{1}{128} \left| x_2\right|  \left| x_1+x_2\right|
   -\frac{1}{16} \left| x_1\right|  \left| x_2-x_3\right|
   -\frac{5}{128} \left| x_2\right|  \left| x_2-x_3\right| \nonumber \\
&-& \frac{5}{128} \left| x_2-x_3\right|  \left| x_3\right|
   -\frac{1}{128} \left| x_2\right|  \left| x_2+x_3\right|
   -\frac{1}{128} \left| x_3\right|  \left| x_2+x_3\right|\nonumber \\
&+& \frac{1}{8} \left| x_1\right|  \left| x_3-x_4\right|
   +\frac{1}{8} \left| x_2\right|  \left| x_3-x_4\right|
   -\frac{1}{16} \left| x_2-x_3\right|  \left| x_3-x_4\right| \nonumber \\
&+& \frac{5}{64} \left| x_3\right|  \left| x_3-x_4\right|
   -\frac{1}{16} \left| x_2-x_3\right|  \left| x_4\right|
   +\frac{5}{64} \left| x_3-x_4\right|  \left| x_4\right| \nonumber \\
&+& \frac{1}{64} \left| x_3\right|  \left| x_3+x_4\right|
   +\frac{1}{64} \left| x_4\right|  \left| x_3+x_4\right|
   -\frac{x_2 \left| x_1\right|  \left| x_1+x_2\right| }{128x_1} \nonumber \\
&+& \frac{x_3 \left| x_2\right|  \left| x_2-x_3\right| }{128x_2}
   -\frac{x_3 \left| x_2\right|  \left| x_2+x_3\right| }{128
   x_2}-\frac{x_4 \left| x_3\right|  \left| x_3-x_4\right| }{64x_3} \nonumber \\
&+& \frac{x_4 \left| x_3\right|  \left| x_3+x_4\right| }{64
   x_3}-\frac{x_1 \left| x_2\right|  \left| x_1+x_2\right| }{128
   x_2}+\frac{x_2 \left| x_2-x_3\right|  \left| x_3\right| }{128 x_3} \nonumber \\
&-& \frac{x_2 \left| x_3\right|  \left| x_2+x_3\right| }{128
   x_3}-\frac{x_3 \left| x_3-x_4\right|  \left| x_4\right| }{64
   x_4}+\frac{x_3 \left| x_4\right|  \left| x_3+x_4\right| }{64
   x_4} \nonumber
\end{eqnarray}

It is easy to see that the infrared divergent term in the above expression,  proportional to $1/\beta$, identically
vanishes, appropriately relabeling the variables:
\begin{eqnarray}
&& \int dx_1 dx_2 dx_3 dx_4  V(x_1) V(x_2) V(x_3) V(x_4)
 \left[ \left| x_1-x_2\right| +\left| x_2-x_3\right| -2 \left|x_3-x_4\right| \right] \nonumber \\
&\rightarrow&
 \int dx_1 dx_2 dx_3 dx_4  V(x_1) V(x_2) V(x_3) V(x_4)
 \left[ \left| x_1-x_2\right| +\left| x_1-x_2\right| -2 \left|x_1-x_2\right| \right] = 0 \nonumber
\end{eqnarray}

Let us now consider the second term, which, after a suitable relabeling reads
\begin{eqnarray}
&& \lambda^4 \int dx_1 dx_2 dx_3 dx_4  V(x_1) V(x_2) V(x_3) V(x_4) \frac{\left(-3 x_1^2+6 x_2 x_1-8 x_2^2-x_3^2+4 x_4^2+10 x_2 x_3-8 x_3 x_4\right)}{64} \nonumber \\
&=&  \lambda^4 \int dx_1 dx_2 dx_3 dx_4  V(x_1) V(x_2) V(x_3) V(x_4) \frac{ x_1 \left( x_2-x_1\right)}{8} \nonumber \\
&=&  \lambda^4 \int dx_1 dx_2 dx_3 dx_4  V(x_1) V(x_2) V(x_3) V(x_4) \frac{ \left( x_2-x_1\right)^2}{16}\nonumber
\end{eqnarray}
where the last line has been obtained upon symmetrization with respect to the variables $x_1$ and $x_2$.

The simplification of the last term requires a bit more of work; the key observation is that, since $V(x_1) V(x_2) V(x_3) V(x_4)$
is completely symmetric in the variables $x_1$,$x_2$,$x_3$ and $x_4$, only the completely symmetric part of
$\mathcal{F}(x_1,x_2,x_3,x_4)$ can contribute.

Upon symmetrization we obtain
\begin{eqnarray}
\mathcal{F}^{({\rm sym})}(x_1,x_2,x_3,x_4) &=& -\frac{1}{96} \left| x_1-x_2\right|  \left| x_1-x_3\right|
   -\frac{1}{96} \left| x_2-x_3\right|  \left| x_1-x_3\right|
   -\frac{1}{96} \left| x_1-x_4\right|  \left| x_1-x_3\right| \nonumber \\
   &-& \frac{1}{48} \left| x_2-x_4\right|  \left| x_1-x_3\right|
   -\frac{1}{96} \left| x_3-x_4\right|  \left| x_1-x_3\right|
   -\frac{1}{96} \left| x_1-x_2\right|  \left| x_2-x_3\right|\nonumber \\
   &-& \frac{1}{96} \left| x_1-x_2\right|  \left| x_1-x_4\right|
   -\frac{1}{48} \left| x_2-x_3\right|  \left| x_1-x_4\right|
   -\frac{1}{96} \left| x_1-x_2\right|  \left| x_2-x_4\right| \nonumber \\
   &-& \frac{1}{96} \left| x_2-x_3\right|  \left| x_2-x_4\right|
   -\frac{1}{96} \left| x_1-x_4\right|  \left| x_2-x_4\right|
   -\frac{1}{48} \left| x_1-x_2\right|  \left| x_3-x_4\right| \nonumber \\
   &-& \frac{1}{96} \left| x_2-x_3\right|  \left| x_3-x_4\right|
   -\frac{1}{96} \left| x_1-x_4\right|  \left| x_3-x_4\right|
   -\frac{1}{96} \left| x_2-x_4\right|  \left| x_3-x_4\right|\nonumber
\end{eqnarray}

With a suitable relabeling it is possible to reduce $\mathcal{F}^{({\rm sym})}(x_1,x_2,x_3,x_4)$ to a simpler form:
\begin{eqnarray}
\mathcal{F}^{({\rm sym})}(x_1,x_2,x_3,x_4) &\rightarrow& -\frac{1}{8} \left| x_1-x_2\right| \left| x_2-x_3\right|
   -\frac{1}{16} \left| x_1-x_2\right|  \left| x_3-x_4\right|\nonumber
\end{eqnarray}

Combining the contributions above one finally obtains the expression for the fourth order
\begin{eqnarray}
E_0^{(4)} &=& - \frac{\lambda^4}{16} \left(\int V(x_1) dx_1  \right)^2 \left( \int V(x_2) | x_2- x_3|^2 V(x_2) dx_2 dx_3  \right) \nonumber \\
&-& \frac{\lambda^4}{8}  \left(\int V(x_1) dx_1  \right) \left( \int V(x_2) | x_2- x_3| V(x_3) | x_3- x_4| V(x_4) dx_2 dx_3 dx_4  \right) \nonumber \\
&-& \frac{\lambda^4}{16} \left( \int V(x_1) | x_1- x_2| V(x_2) dx_1 dx_2  \right)^2
\end{eqnarray}
which agrees  with the expression calculated by Patil~\footnote{Note that the different convention that we are using for the kinetic term. }.

\subsection{Fifth order}
\label{fifth}

The calculation of the higher order contributions is performed in the similar way as for the fourth order; in
the case of the fifth order contribution the expression contains potentially infrared divergent terms
of order $1/\beta^3$, $1/\beta^2$ and $1/\beta$.
Upon symmetrization and suitable relabeling of the integration variables  one can show that each
of these contributions identically vanishes, as expected.

As a result, only the term of order $\beta^0$ survives and, upon simplification, it takes the form
\begin{eqnarray}
E_0^{(5)} &=& - \frac{\lambda^5}{96} \left(\int V(x_1) dx_1  \right)^3 \left( \int V(x_2) | x_2- x_3|^3 V(x_2) dx_2 dx_3  \right) \nonumber \\
&-& \frac{\lambda^5}{16} \left(\int V(x_1) dx_1  \right)^2 \left( \int V(x_2) | x_2- x_3| V(x_3) | x_3- x_4|^2 V(x_4) dx_2 dx_3 dx_4   \right) \nonumber \\
&-& \frac{\lambda^5}{16} \left(\int V(x_1) dx_1  \right) \left( \int V(x_2) | x_2- x_3| V(x_3) | x_3- x_4| V(x_4) |x_4-x_5| V(x_5) dx_2 dx_3 dx_4 dx_5   \right) \nonumber \\
&-& \frac{\lambda^5}{16} \left(\int V(x_1) dx_1  \right) \left( \int V(x_2) | x_2- x_3| V(x_3)  dx_2 dx_3   \right) \left( \int V(x_4) | x_4- x_5|^2 V(x_5)  dx_4 dx_5   \right) \nonumber \\
&-& \frac{\lambda^5}{16} \left( \int V(x_1) | x_1- x_2| V(x_2)  dx_1 dx_2   \right) \left( \int V(x_3) | x_3- x_4| V(x_4)| x_4- x_5| V(x_5)  dx_3 dx_4 dx_5   \right)
\end{eqnarray}
that agrees with the fifth order contribution calculated by Patil.

\subsection{Sixth order}
\label{sixth}

The calculation of the sixth order contribution is considerably more involved than the fifth order, although it
can be performed along the same lines. In this case, there are divergent contributions of order
$1/\beta^4$, $1/\beta^3$, $1/\beta^2$ and $1/\beta$.
Once again, one finds that each of these contributions identically vanishes when a symmetrization of the
integrands and a suitable relabeling of the integration variable is carried out.

After performing all the algebra, the simplest form that we have obtained for the sixth order term is
\begin{eqnarray}
E_0^{(6)} &=& \lambda^6
\int dx_1 dx_2 dx_3 dx_4 V(x_1)V(x_2) V(x_3) V(x_4) \nonumber \\
&\times&
\left( -\frac{x_1^4}{96}+\frac{1}{24} x_2 x_1^3-\frac{5}{64} x_2^2 x_1^2+\frac{3}{32} x_2 x_3
   x_1^2-\frac{3}{64} x_2 x_3 x_4 x_1 \right) \nonumber \\
&\times&    \left(\int V(x_5) dx_5  \right)^2  + \lambda^6
\int dx_1 dx_2 dx_3 dx_4 dx_5 dx_6 V(x_1)V(x_2) V(x_3) V(x_4) V(x_5) V(x_6) \nonumber \\
&\times& \left[ \left(-\frac{1}{48} \left(x_1-x_2\right)^2-\frac{1}{32} \left(x_2-x_3\right) \left(x_1-x_2\right)
-\frac{1}{32} \left(x_3-x_4\right) \left(x_1-x_2\right) \right. \right.  \nonumber \\
&-& \left. \left. \frac{1}{48} \left(x_4-x_5\right) \left(x_1-x_2\right)
-\frac{1}{96} \left(x_5-x_6\right) \left(x_1-x_2\right)-\frac{1}{48}
   \left(x_2-x_3\right)^2-\frac{1}{32} \left(x_3-x_4\right)^2 \right. \right.  \nonumber \\
&-& \left. \left. \frac{1}{24} \left(x_4-x_5\right)^2-\frac{1}{32} \left(x_5-x_6\right)^2
-\frac{1}{32} \left(x_2-x_3\right) \left(x_3-x_4\right) \right. \right. \nonumber \\
&-& \left. \left. \frac{1}{48} \left(x_2-x_3\right) \left(x_4-x_5\right)-\frac{1}{24} \left(x_3-x_4\right) \left(x_4-x_5\right)
-\frac{1}{96} \left(x_2-x_3\right) \left(x_5-x_6\right) \right. \right. \nonumber \\
&-& \left. \left. \frac{1}{48} \left(x_3-x_4\right) \left(x_5-x_6\right)-\frac{1}{24} \left(x_4-x_5\right) \left(x_5-x_6\right)\right)
   \left|x_1-x_2\right|  \left| x_2-x_3\right| \right. \nonumber \\
&+& \left.  \left(-\frac{1}{32} \left(x_1-x_2\right)^2-\frac{3}{64} \left(x_2-x_3\right) \left(x_1-x_2\right)-\frac{5}{128} \left(x_3-x_4\right) \left(x_1-x_2\right) \right.\right. \nonumber \\
&-& \left. \left. \frac{1}{32} \left(x_4-x_5\right) \left(x_1-x_2\right)-\frac{1}{64} \left(x_5-x_6\right) \left(x_1-x_2\right)-\frac{3}{64} \left(x_2-x_3\right)^2
\right.\right. \nonumber \\
&-& \left. \left.  \frac{1}{16} \left(x_3-x_4\right)^2-\frac{1}{16} \left(x_4-x_5\right)^2-\frac{3}{64} \left(x_5-x_6\right)^2 \right.\right. \nonumber \\
&-& \left. \left. \frac{5}{64} \left(x_2-x_3\right) \left(x_3-x_4\right)-\frac{1}{16} \left(x_2-x_3\right) \left(x_4-x_5\right)
-\frac{3}{32} \left(x_3-x_4\right) \left(x_4-x_5\right) \right.\right. \nonumber \\
&-& \left. \left. \frac{1}{32} \left(x_2-x_3\right) \left(x_5-x_6\right)-\frac{3}{64}
   \left(x_3-x_4\right) \left(x_5-x_6\right) \right.\right. \nonumber \\
&-& \left. \left. \frac{1}{16} \left(x_4-x_5\right) \left(x_5-x_6\right)\right) \left| x_1-x_2\right|  \left| x_3-x_4\right|
\right] \nonumber \\
&+& \lambda^6
\int dx_1 dx_2 dx_3 dx_4 dx_5 dx_6 V(x_1)V(x_2)V(x_3) V(x_4) V(x_5) V(x_6)\nonumber \\
&\times& \left( -\frac{1}{32} \left| x_1-x_2\right|  \left| x_2-x_3\right|
   \left| x_3-x_4\right|  \left| x_4-x_5\right| -\frac{1}{64}
   \left| x_1-x_2\right|  \left| x_2-x_3\right|  \left|
   x_5-x_6\right|  \left| x_4-x_5\right|  \right. \nonumber \\
   &-&  \left. \frac{1}{32} \left|
   x_1-x_2\right|  \left| x_2-x_3\right|  \left|
   x_3-x_4\right|  \left| x_5-x_6\right|\right) \ .
\end{eqnarray}

\section{Applications}
\label{sec:App}

We have applied this expression to two exactly solvable problems, the  finite square well~\footnote{We assume $v_0>0$.}
\begin{eqnarray}
v(x) = \left\{
\begin{array}{ccc}
0 & , & |x|>a \\
-v_0 & , & |x|< a
\end{array}
\right. \nonumber
\end{eqnarray}
and the P\"osch-Teller potential\cite{F99}
\begin{eqnarray}
v(x) = -\frac{v_0}{\cosh^2 x} \nonumber
\end{eqnarray}

In both cases we have reproduced the results obtained from the exact result, expanding to order $\lambda^6$.

For the finite square well we obtain
\begin{eqnarray}
\epsilon &=& -v_0^2+\frac{4 v_0^3}{3}-\frac{92 v_0^4}{45}+
\frac{1072 v_0^5}{315} - \frac{84752 v_0^6}{14175} + O\left(v_0^{7}\right) \nonumber
\end{eqnarray}
whereas for the P\"osch-Teller potential we obtain
\begin{eqnarray}
\epsilon &=& -v_0^2-2 v_0^3-5 v_0^4-14 v_0^5-42 v_0^6+ O\left(v_0^{7}\right) \nonumber
\end{eqnarray}

We have also applied this formula to the case of a gaussian well
\begin{eqnarray}
V(x) = -v_0 e^{-x^2} \nonumber
\end{eqnarray}

To order $v_0^6$ we have
\begin{eqnarray}
\epsilon &=&  -\frac{1}{4} \pi  v_0^2  + \frac{\pi  v_0^3}{2 \sqrt{2}} - \left(\frac{\pi }{8}+\frac{\sqrt{3} \pi }{8}+\frac{\pi ^2}{12}\right) v_0^4
+ \left(\frac{7 \pi }{96}+\frac{1}{8} \sqrt{\frac{3}{2}} \pi +\frac{3 \pi ^2}{8 \sqrt{2}} + \int_{-\infty}^\infty F(x) dx
\right) v_0^5 \nonumber \\
&+& \left(-\frac{3 \pi }{64}-\frac{7 \pi }{96
   \sqrt{2}}-\frac{7 \pi }{96 \sqrt{5}}-\frac{5 \pi
   ^2}{16}-\frac{\pi ^2}{64 \sqrt{3}}-\frac{7
   \sqrt{3} \pi ^2}{64}-\frac{2 \pi ^3}{45} + \int_{-\infty}^\infty G(x) dx \right) v_0^6  + O(v_0^7)\nonumber \\
   &\approx& -0.785398 v_0^2 + 1.11072 v_0^3 -1.89534 v_0^4 + 3.56727 v_0^5 -7.1374 v_0^6 + O(v_0^7)
   \label{pert}
\end{eqnarray}
where
\begin{eqnarray}
F(x) &\equiv& \frac{\pi ^{3/2} e^{-2 x^2} }{128} \left(e^{x^2} x (2 {\rm erf}(x)-1) \left(4 \sqrt{2}
   x {\rm erf}\left(\sqrt{2} x\right)-\sqrt{\pi } {\rm erf}(x)^2\right)-2
   {\rm erf}(x)^2\right) \nonumber \\
G(x) &\equiv& \frac{\pi ^2 e^{-x^2} x {\rm erf}(x)^3}{64 \sqrt{2}}+\frac{\pi ^2 e^{-x^2} x {\rm erf}\left(\sqrt{2} x\right)
   {\rm erf}(x)^2}{32 \sqrt{2}} + \frac{1}{64} \pi^{3/2} e^{-3 x^2} {\rm erf}(x)^2 \nonumber \\
   &+& \frac{\pi
   ^{3/2} e^{-2 x^2} {\rm erf}(x)^2}{64
   \sqrt{2}}-\frac{1}{16} \pi ^{3/2} e^{-x^2} x^2
   {\rm erf}\left(\sqrt{2} x\right)
   {\rm erf}(x)-\frac{1}{16} \pi ^{3/2} e^{-x^2}
   x^2 {\rm erf}\left(\sqrt{2} x\right)^2 \nonumber
\end{eqnarray}

In this case we do not dispose of the exact result to compare
with, since the problem in not exactly solvable. However, we can
easily apply a Pad\'e approximant to the perturbative expression
above, after having singled out the asymptotic behavior for $v_0
\rightarrow \infty$; the resummed expression reads
\begin{eqnarray}
\tilde{\epsilon} = -v_0 +\frac{v_0 +2.60002 v_0 ^2+1.2553 v_0^3}{1+3.38542 v_0 +2.80348 v_0 ^2+0.336931 v_0 ^3}
\label{Pade}
\end{eqnarray}

\begin{figure}
\begin{center}
\bigskip\bigskip\bigskip
\includegraphics[width=8cm]{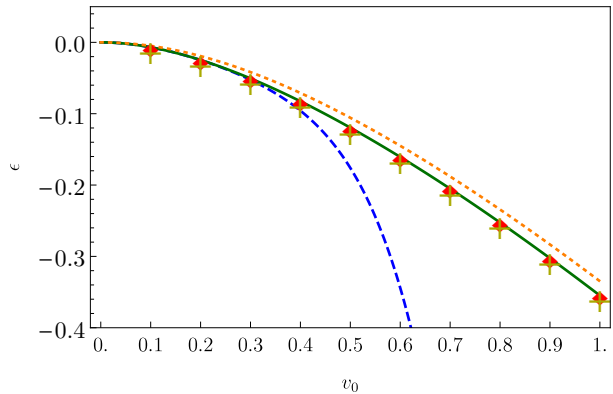}
\caption{Energy of the gaussian as a function of the depth. The
solid green line is the Pad\'e approximant (\ref{Pade}), whereas
the dashed blue line is the perturbative expression of
eq.(\ref{pert}), the orange dotted line and the red rhombi are the
variational energy obtained with the wave functions (\ref{var1})
and (\ref{var2}) respectively. Finally the crosses are the precise
results obtained with the Wronskian method.} \label{Fig_1}
\end{center}
\end{figure}

In Fig.~\ref{Fig_1} we compare the energy estimated with Eq.(\ref{Pade}) (solid green line), with the perturbative expression to sixth order
of Eq.~(\ref{pert}) (dashed blue line), and with two variational estimates obtained with the trial wave functions
\begin{eqnarray}
\psi(x) = e^{-\alpha x^2}
\label{var1}
\end{eqnarray}
and
\begin{eqnarray}
\psi(x) = e^{-\alpha \sqrt{\beta^2+x^2}}
\label{var2}
\end{eqnarray}
which are respectively represented by the orange dotted line and
by the red rhombi. Finally the crosses are accurate numerical
results obtained by means of the Wronskian method\cite{F11}.

The second wave function has the correct decay at $|x| \rightarrow
\infty$: we observe an excellent agreement between the variational
energy obtained in this case, minimizing with respect to the
parameters $\alpha$ and $\beta$, and the energy obtained in
eq.~(\ref{Pade}), using the Pad\'e approximant of the perturbative
expression to sixth order. Notice that the Pad\'e approximant is
completely analytical and does not require to introduce additional
parameters.

\section{Conclusions}
\label{concl}

The calculations contained in the present paper on one hand
confirm the soundness of the method originally developed by Gat
and Rosenstein, reproducing the perturbative corrections to the
energy of the ground state in a weak short range finite well,
previously calculated with different techniques, while on another
hand they provide the contribution to sixth order, which had not
been calculated before. In our view, this method has the
attractive feature of allowing to apply the usual
Rayleigh-Schr\"odinger perturbation theory to a problem with a
mixed (discrete-continuum) spectrum, which is intractable if one
uses the free hamiltonian as the unperturbed hamiltonian
$\hat{H}_0$, due to the presence of infrared singularities. In the
scheme of Gat and Rosenstein, these singularities manifest as
terms proportional to inverse powers of $\beta$ (the strength of
the artificial delta potential) and turn out to exactly vanish at
each perturbative order. We have applied the formula to sixth
order to two exactly solvable examples, reproducing the results
obtained from the exact expressions, upon expansion in the
perturbative parameter. For the case of the (not-exactly solvable)
gaussian well, we have compared the analytic expression obtained
applying a Pad\'e approximant to the perturbative results with the
precise numerical results obtained variationally and with the
method of Ref.~\cite{F11}, observing an excellent agreement.

\section*{Acknowledgements}
The research of P.A. was supported by the Sistema Nacional de Investigadores (M\'exico).

\appendix

\section{Green's functions}
\label{AppA}

In analogy with ref.~\cite{Amore16} we define the operator
\begin{eqnarray}
\hat{\Omega}_\gamma \equiv \int_0^\infty \frac{dp}{2\pi} \frac{|p\rangle\langle p |}{\epsilon_p -\epsilon_0 + \gamma}
\end{eqnarray}
where $\epsilon_p = p^2$ and $\epsilon_0 = -\beta^2$. The Dirac bra-ket notation is used for the eigenstates of $\hat{H}_0$
belonging to the continuum.

In terms of this operator we define the Green's function
\begin{eqnarray}
\mathcal{G}_\gamma(x_1,x_2)  &\equiv& \langle x_1 | \hat{\Omega}_\gamma | x_2\rangle \nonumber \\
&=& \theta \left(x_1\right) \theta \left(x_1-x_2\right) \theta \left(x_2\right) \nonumber \\
&\times& \frac{
   e^{\left(x_1+x_2\right) (-(\beta +\Gamma ))} \left(e^{\beta  \left(x_1+x_2\right)}
   \left((\beta +1) \Gamma +\gamma  \left(e^{2 \Gamma  x_2}-1\right)\right)-2 \beta
   \Gamma  e^{\Gamma  \left(x_1+x_2\right)}\right)}{2 \gamma  \sqrt{\Gamma }} \nonumber \\
   &+& \frac{\theta \left(-x_1\right) \theta \left(x_1-x_2\right) \theta \left(-x_2\right)
   \left(e^{\Gamma  \left(x_1+x_2\right)} (\beta  \Gamma -\gamma +\Gamma )-2 \beta
   \Gamma  e^{\beta  \left(x_1+x_2\right)}+\gamma  e^{\Gamma
   \left(x_2-x_1\right)}\right)}{2 \gamma  \sqrt{\Gamma }} \nonumber \\
   &+& \frac{\sqrt{\Gamma } \theta \left(x_1-x_2\right) \theta \left(x_1\right) \theta \left(-x_2\right) e^{x_1 (-(\beta
   +\Gamma ))} \left(\beta  \left(e^{\beta  x_1+\Gamma  x_2}-2 e^{\beta  x_2+\Gamma
   x_1}\right)+e^{\beta  x_1+\Gamma  x_2}\right)}{2 \gamma } \nonumber \\
&+& \theta \left(x_1\right) \theta \left(x_2\right) \theta \left(x_2-x_1\right) \nonumber \\
&\times& \frac{
   e^{\left(x_1+x_2\right) (-(\beta +\Gamma ))} \left(\gamma  e^{\beta
   \left(x_1+x_2\right)+2 \Gamma  x_1}+e^{\beta  \left(x_1+x_2\right)} (\beta  \Gamma
   -\gamma +\Gamma )-2 \beta  \Gamma  e^{\Gamma  \left(x_1+x_2\right)}\right)}{2 \gamma
   \sqrt{\Gamma }} \nonumber \\
   &+& \frac{\theta \left(-x_1\right) \theta \left(-x_2\right) \theta \left(x_2-x_1\right)
   \left(e^{\Gamma  \left(x_1+x_2\right)} (\beta  \Gamma -\gamma +\Gamma )-2 \beta
   \Gamma  e^{\beta  \left(x_1+x_2\right)}+\gamma  e^{\Gamma
   \left(x_1-x_2\right)}\right)}{2 \gamma  \sqrt{\Gamma }} \nonumber \\
   &+& \frac{\sqrt{\Gamma } \theta \left(-x_1\right) \theta \left(x_2\right) \theta \left(x_2-x_1\right) \left(\beta
   \left(e^{\Gamma  \left(x_1-x_2\right)}-2 e^{\beta
   \left(x_1-x_2\right)}\right)+e^{\Gamma  \left(x_1-x_2\right)}\right)}{2 \gamma }
\end{eqnarray}
where
\begin{eqnarray}
\Gamma \equiv \sqrt{\beta^2+\gamma}
\end{eqnarray}

We have
\begin{eqnarray}
\mathcal{G}_\gamma(x_1,x_2)  = \sum_{\ell=0}^\infty (-1)^\ell \mathcal{G}^{(\ell)}(x_1,x_2)
\end{eqnarray}
where
\begin{eqnarray}
\mathcal{G}^{(\ell)}(x_1,x_2) = \langle x_1 | \hat{\Omega}^{\ell+1} | x_2\rangle
\end{eqnarray}
are the Green's functions needed in the application of the perturbative method.

The explicit expressions for the first few Green's functions are
\begin{eqnarray}
\mathcal{G}^{(0)}(x_1,x_2) &=& \frac{1}{4 \beta } + \frac{1}{4} \left(-\left| x_1\right| -2 \left| x_1-x_2\right| -\left| x_2\right| \right) + O(\beta) \nonumber \\
\mathcal{G}^{(1)}(x_1,x_2) &=& \frac{1}{16 \beta ^3}-\frac{\left| x_1\right|
   +\left| x_2\right| }{16 \beta ^2}+\frac{2
   \left| x_1\right|  \left| x_2\right|
   -3 x_1^2+8 x_1 x_2-3 x_2^2}{32 \beta } \nonumber \\
   &+& \frac{1}{96} \left(8
   \left| x_1-x_2\right|
   \left(x_1-x_2\right)^2+3 \left| x_2\right|
   \left(3 x_1^2+x_2^2\right)+3 \left| x_1\right|
   \left(x_1^2+3 x_2^2\right)\right) + O(\beta) \nonumber \\
\mathcal{G}^{(2)}(x_1,x_2) &=& \frac{1}{32 \beta ^5}
   -\frac{\left| x_1\right| +\left| x_2\right| }{32 \beta ^4}
   -\frac{-2 \left| x_1\right|  \left| x_2\right| +x_1^2-4 x_1 x_2+x_2^2}{64 \beta^3} \nonumber \\
   &+& \frac{\left(\left| x_1\right| +3 \left|x_2\right| \right) x_1^2+\left(3 \left|x_1\right| +\left| x_2\right| \right)
   x_2^2}{192 \beta ^2} \nonumber \\
   &+& \frac{5 x_1^4-24 x_1^3 x_2+30 x_1^2
   x_2^2-24 x_1 x_2^3+5 x_2^4-4 \left| x_1\right|
   \left| x_2\right|  \left(x_1^2+x_2^2\right)}{768
   \beta } \nonumber \\
   &+& \frac{-16 \left| x_1-x_2\right|
   \left(x_1-x_2\right)^4-5 \left| x_2\right|
   \left(5 x_1^4+10 x_1^2 x_2^2+x_2^4\right)-5 \left|
   x_1\right|  \left(x_1^4+10 x_1^2 x_2^2+5
   x_2^4\right)}{3840} + O(\beta) \nonumber \\
\mathcal{G}^{(3)}(x_1,x_2) &=& \frac{5}{256 \beta ^7}
   -\frac{5 \left(\left|x_1\right| +\left| x_2\right| \right)}{256\beta ^6}
   +\frac{10 \left| x_1\right| \left| x_2\right| -3 x_1^2+16 x_1 x_2-3 x_2^2}{512 \beta ^5} \nonumber \\
   &+& \frac{\left(\left|x_1\right| +3 \left| x_2\right| \right) x_1^2+\left(3 \left| x_1\right| +\left|x_2\right| \right) x_2^2}{512 \beta^4} \nonumber \\
   &+& \frac{5 x_1^4-32 x_1^3 x_2+30 x_1^2 x_2^2-32 x_1 x_2^3+5 x_2^4-12 \left| x_1\right|  \left|x_2\right|
   \left(x_1^2+x_2^2\right)}{6144 \beta^3} \nonumber \\
   &-& \frac{\left(\left| x_1\right| +5 \left|x_2\right| \right) x_1^4+10 \left(\left|x_1\right| +\left| x_2\right| \right) x_1^2 x_2^2+\left(5 \left| x_1\right| +\left| x_2\right| \right) x_2^4}{6144 \beta^2} \nonumber \\
   &+& \frac{1}{36864 \beta } \left[ -7 x_1^6+48 x_1^5 x_2-105 x_1^4 x_2^2+160 x_1^3 x_2^3-105 x_1^2 x_2^4+48 x_1 x_2^5 \right. \nonumber \\
     &-& \left. 7 x_2^6+2 \left|x_1 x_2\right|
     \left(3 x_1^2+x_2^2\right) \left(x_1^2+3 x_2^2\right)\right]
   \nonumber \\
   &+& \frac{1}{1290240} \left[ 128 \left| x_1-x_2\right|
     \left(x_1-x_2\right)^6 \right. \nonumber \\
     &+& \left. 35 \left| x_2\right|
   \left(7 x_1^6+35 x_1^4 x_2^2+21 x_1^2 x_2^4+x_2^6\right)+35
   \left| x_1\right|  \left(x_1^6+21 x_1^4 x_2^2+35
   x_1^2 x_2^4+7 x_2^6\right) \right]
   \nonumber \\
   &+& O(\beta)
\end{eqnarray}


\end{document}